\documentclass{iaus}
\usepackage{graphicx}

\title[Migration] 
{Orbital Migration Models under Test}

\author[Wilhelm Kley]   
{Wilhelm Kley}
\affiliation{Institut f\"ur Astronomie \& Astrophysik \\
  Universit\"at T\"ubingen, Morgenstelle 10, 72076 T\"ubingen, Germany  \\
 email: {\tt wilhelm.kley@uni-tuebingen.de}}
\pubyear{2010}
\volume{276}  
\pagerange{119--126}
\jname{The Astrophysics of Planetary Systems: \\
   Formation, Structure, and Dynamical Evolution}
\editors{A.~Sozzetti, M.G.~Lattanzi \& A.P.~Boss, eds.}
\begin{document}
\maketitle
\begin{abstract}
Planet-disk interaction predicts a change in the orbital elements of an embedded planet.
Through linear and fully hydrodynamical studies it has been found that migration is typically directed
inwards. Hence, this migration process gives natural explanation for the presence  
of the 'hot' planets orbiting close to the parent star, and
it plays a mayor role in explaining the formation of resonant planetary systems.

However, standard migration models for locally isothermal disks indicate a too rapid inward migration for small
mass planets, and a large number of massive planets are found very far away from the star.
Recent studies, including more complete disk physics, have opened up new paths
to slow down or even reverse migration. The new findings on migration are discussed and connected to the
observational properties of planetary systems.

\keywords{planetary systems: formation, accretion disks, hydrodynamics}
\end{abstract}

\firstsection 
\section{Introduction}

The orbital elements of the observed extrasolar planets are distinctly different from the solar system.
The major solar system planets have a nearly coplanar configuration and orbits with small eccentricity.
In contrast, the exoplanet population displays large eccentricities, and many planets orbit their host
star on very tight orbits. Recently, it has been discovered that inclined and retrograde orbits are quite frequent as well,
at least for close-in planets. Historically, it was exactly the relatively 'calm' dynamical structure
of the solar system that led to the hypothesis that planets form in protoplanetary disks.
Within the framework of the sequential accretion scenario planet formation proceeds along a series of
substeps, growing from small dust particles all the way to the giant gaseous planets.
The discovery of the hot planets, which could not have been formed in-situ due to the hot temperatures
and limited mass reservoir, gave rise to the exploration of dynamical processes 
that are able to change the location of planets in the disk, accompanying the regular
formation process.

In the context of moons embedded in the ring system of Saturn it had been noted that 
disk-satellite interaction can alter the orbital elements of the perturbing moon,
in particular its semi-major axis (\cite{1980ApJ...241..425G}).
Later it was recognized that via the very same process, operating between a young embedded planet and the  
protoplanetary disk, it is possible to bring a planet that has formed at large distances from the star to its
close proximity (\cite{1986Icar...67..164W}). 
Hence, this migration process has provided a natural explanation for the population of hot planets,
and their mere existence has been considered as evidence for the migration process.
At the same time it was noted that planet-disk interaction may lead to eccentricity 
as well as inclination damping (\cite{1988Icar...73..330W, 1994Icar..110...95W}).

Another indication for a planetary migration process comes from the high fraction (nearly 20\%) of 
configurations in a low order mean-motion resonance, within the whole sample of multi-planet systems.
As the direct formation of such systems seems unlikely, only a dissipative process that changes the
energy (semi-major axis) is able to bring planets from their initial non-resonant configuration
into resonance. Since resonant capture excites the planetary eccentricities typically to large values 
in contradiction to the archetypical system GJ~876, it has been inferred that planet-disk interaction
should lead to eccentricity damping (\cite{2002ApJ...567..596L}).

However, population synthesis models strongly indicated that the standard migration scenario
yields very rapid inward migration rates that appear to be in disagreement with the observations.
Additionally, recent observations of close-in planets on eccentric and inclined orbits have questioned the general
validity of the migration paradigm to form them.
In this review, I will first explain the basic mechanism of migration, present new findings on the migration rate, and
then discuss its applicability with respect to the overall planet formation process.

\section{Origin of migration}

An embedded object disturbs the ambient disk dynamically in two important ways: First it divides the disk into an inner and outer
disk separated by a coorbital (horse shoe) region. Secondly, the propagating sound waves that are sheared
out by the Keplerian differential rotation generate density waves in the form of spiral arms in the disk.
The created structures in the perturbed coorbital region and in the spiral arms
back-react on the planet and cause a change in its semi-major axis.
Thus, physically speaking, planetary migration is caused by the effect of spiral arms and corotation region.
Let us discuss these effects in turn.

\underline {Spiral arms}: \, 
To put it simple, the spiral arms can be considered as density enhancements in the disk that 'pull' gravitationally on the planet.
This gives rise to so called Lindblad torques that change the planet's angular momentum. For circular orbits the disk torque
exerted on the planet is directly a measure of the speed and direction of migration.
The inner spiral forms a leading wave that causes a positive torque, while the outer wave generates a negative
contribution. The combined effects of both spirals determine then the sign and magnitude of the total torque.
A positive total torque will add angular momentum to the planet and cause outward migration. On the other hand, a
negative torque will induce inward migration.
It turns out that under typical physical disk conditions the contributions of the inner and outer spiral
arm are comparable in magnitude. However, the effect of the outer spiral quite generally
wins over the inner one causing the planet to migrate inward. 

\underline {Corotation region}: \, 
As viewed in the corotating frame, material within the corotation region performs so called horseshoe orbits. Here, the
gas particles upon approaching the planet at the two ends of the horseshoe
are periodically shifted from an orbit with a semi-major axis slightly
larger than the planetary one to an orbit with slightly smaller value, and vice versa. Hence, at each close
approach with the planet there is an exchange of angular momentum between (coorbital) disk material and the planet.
The total corotation torque is then obtained by adding the contributions from both ends of the horseshoe. To obtain
a net, non-zero torque requires non-vanishing radial gradients of vortensity and entropy across the corotation region
(\cite{2008ApJ...672.1054B}).
For an ideal gas without friction or heat diffusion mixing effects within the horseshoe tend to flatten out these
gradients yielding a vanishing corotation torque, or so called torque saturation.

\subsection{Type-I migration}
Small mass planets do not alter the global disk structure significantly, in particular they do not open a gap within
disk. Hence, the combined effect of Lindblad and corotation torques can be calculated for small planetary masses using
a linear analysis. The outcome of such linear, no-gap studies has been termed type-I planet migration.
Due to the complexity of considering heat generation and transport in disks these studies have
relied nearly exclusively on simplified, locally isothermal disk models. 
Here, the temperature is assumed to be independent of height and is given by a pre-described function of radius, $T=T(r)$.
Typically, it is assumed that the relative scale height $H$ of the disk is a constant, $H/r=const.$, yielding $T \propto r^{-1}$.
The total torque $\Gamma_{tot}$ is given as the sum of Lindblad and corotation torque $\Gamma_{tot} = \Gamma_L + \Gamma_{CR}$. 
The speed of the induced linear, type-I migration scales inversely
with the disk temperature (i.e. disk thickness) as $\propto (H/r)^{-2}$, linear with the planet mass 
$\propto m_p$, and with the disk mass $\propto m_d$.
Linear models have been calculated for flat 2D disks as well as full 3D configurations. The problem of 2D simulations
lies in taking into account approximately the neglected vertical stratification of the disk, which is
typically done through a smoothing of the gravitational potential near the planet. Additional problems arise when considering
radial gradients. Hence, 2D and 3D results may well yield agreeing migration rates at a particular radius but 
opposite dependence on radial gradients. Additionally, non-linear effects may set in already at a planetary mass of 
about 10 earth masses. New full 3D, nested grid locally isothermal hydrodynamic simulations of planet-disk interaction
give very good agreement with previous 3D linear results (\cite{2004ApJ...602..388T}) 
and yield the following form for the total torque for small mass planets below about 10 $M_{Earth}$ (\cite{2010ApJ...724..730D}) 
\begin{equation}
\label{eq:kley-gennaro}
        \Gamma_{tot}  = - (1.36 + 0.62 \alpha_\Sigma + 0.43 \alpha_T) 
       \, \left(\frac{m_{p}}{M_*}\right)^{2} 
       \, \left(\frac{H}{r_p}\right)^{-2}   \, \Sigma_p \, r_p^4 \, \Omega_{p}^2.
\end{equation}
In eq.~(\ref{eq:kley-gennaro}) the index $p$ refers to the planet, $\alpha_\Sigma$ and $\alpha_T$ refer
to the radial variation of density and temperature, such that $\Sigma(r) \propto r^{-\alpha_\Sigma}$
and $T(r) \propto r^{-\alpha_T}$. 

\subsection{Type-II migration}
\begin{figure}[ht]
\begin{center}
\includegraphics[width=0.40\textwidth]{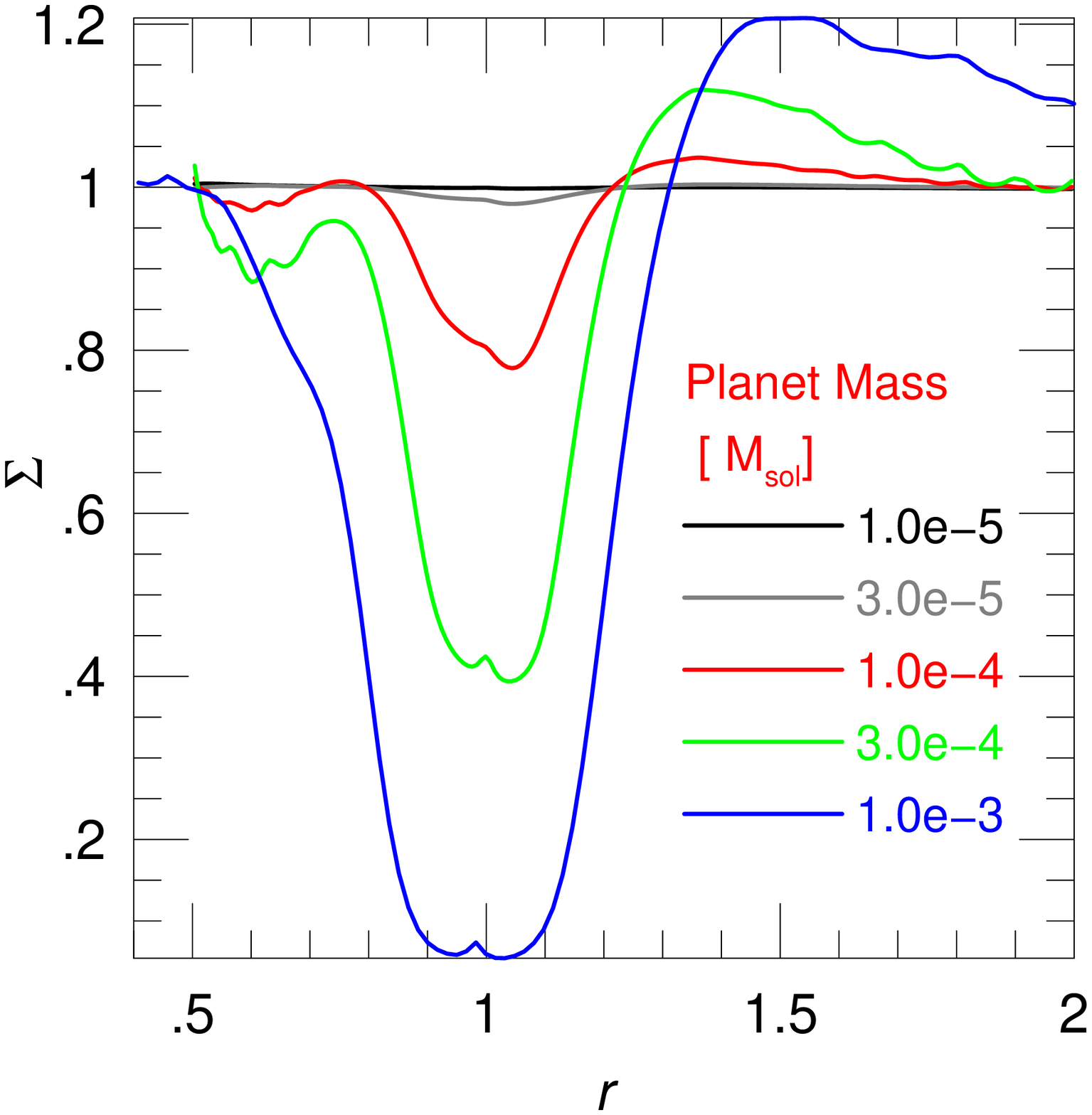} 
$~~$
\includegraphics[width=0.35\textwidth]{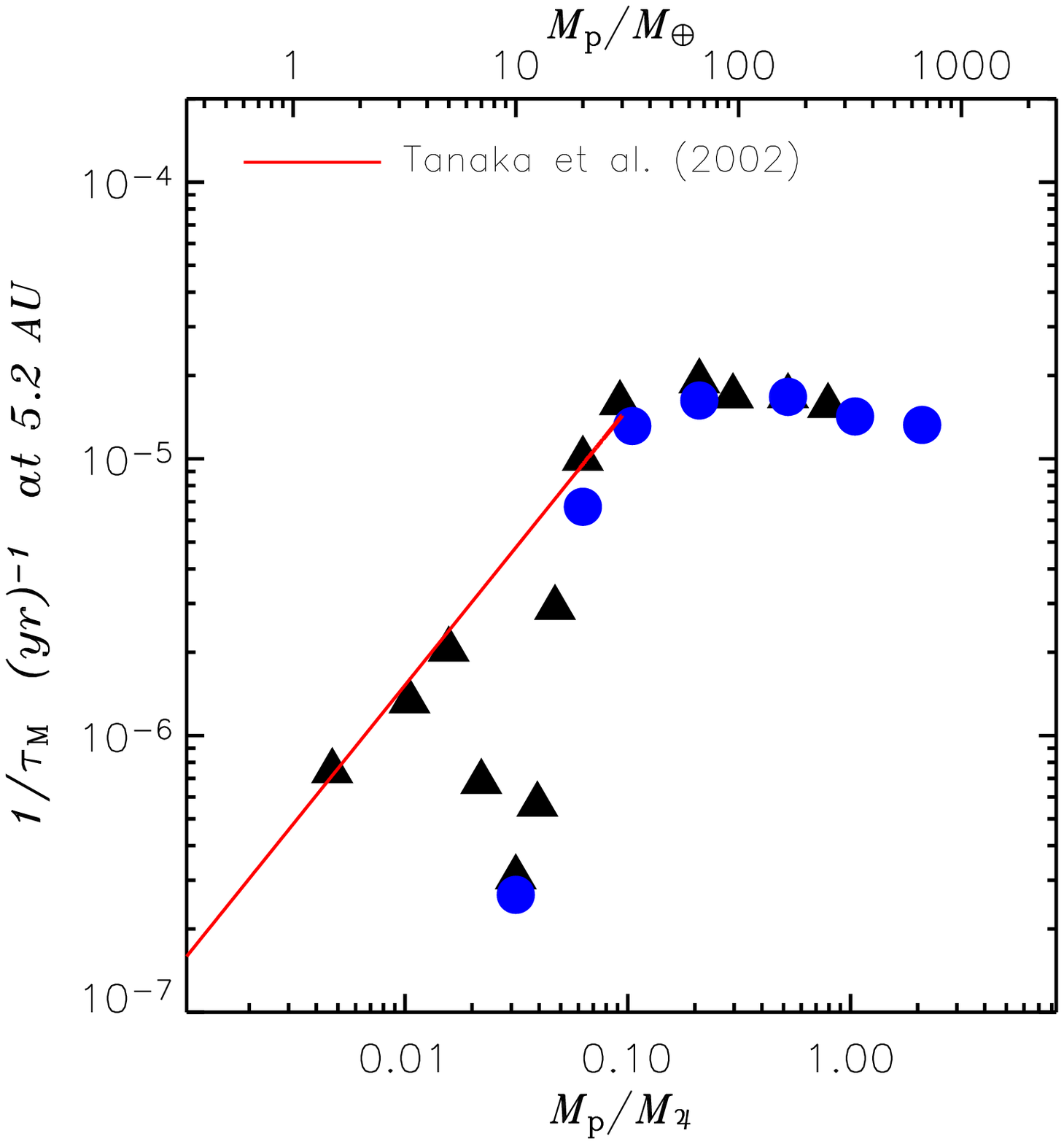} 
 \caption{{\bf Left}: Azimuthally averaged surface density profile of the disk for a variety of planet masses, quoted in solar
 masses. The density structure is obtained for an isothermal disk with $H/r=0.05$ using a constant viscosity, equivalent to
 a $\alpha = 0.004$ at the radius of the planet, as obtained with 2D hydrodynamic simulations.  
 {\bf Right}: The migration rate, quoted in terms of the migration time scale $\tau = a/\dot{a}$ for different planet masses.
 The results refer to 3D nested grid simulations where the symbols denote different grid layouts. The red solid line
 denotes the result of \cite{2002ApJ...565.1257T}.
}
   \label{fig:kley01}
\end{center}
\end{figure}

For larger planet masses, a gap is opened in the disk because the planet transfers angular momentum
to the disk, positive exterior and negative interior to the planet. The depth of the gap that the planet carves out depends
for given disk physics (temperature and viscosity) only on the mass of the planet (Fig.~\ref{fig:kley01}, left panel).
Because the density in the coorbital region is reduced, the corotation torques are strongly affected and are no longer
of any importance for larger planet masses. For very large masses even the Lindblad torques are reduced yielding a slowing down
of the planet (Fig.~\ref{fig:kley01}, right panel). This non-linear regime has been coined the type-II regime of planetary migration;
here the drift of the planet is dominated by the disk's viscous evolution.
The dip in the migration rate at around $m_p = 10 M_{Earth}$ has been discovered by \cite{2006ApJ...652..730M}, and can be
attributed to a change in the flow structure in the vicinity of the planet due to non-linear effects.

\subsection{Migration in radiative disks}

The migration rates obtained through linear analysis as well as fully non-linear hydrodynamical models
have resulted in approximate formulae for the
migration speed $\dot{a}$ of a planet (as quoted above for small mass regime, eq.~\ref{eq:kley-gennaro})
that are frequently used in population synthesis models,
i.e. growth models of planets that include disk evolution and planet migration in parameterized form as well. 
These synthesis models can be used to calculate the final location of many planets in the mass/semi-major axis
diagram and compare the results statistically with the observed distribution. 
Through variation of individual parameter of the model, their relative importance can be estimated.  
The results indicate in particular, that the migration for small mass planets in the type-I regime
is by far too fast to account for the observed distribution, the majority of planets would have been lost to the star,
as shown for example by \cite{2008ApJ...673..487I} and \cite{2009A&A...501.1161M}.
Only a significant reduction in the type-I migration speed gives satisfactory results. 
Suggested remedies included: stochastic migration of a planet in a turbulent disk, migration in inviscid
self-gravitating disks or nonlinear effects.

Here, we shall concentrate on a very simple and straight forward improvement of the models, that represent a
possible solution to the type-I migration problem: the inclusion of more accurate physics. As mentioned above,
past modeling relied nearly exclusively on the simplified locally isothermal models, which has the advantage that
no energy equation has to be considered. Taking more realistic thermodynamics into account requires
the incorporation of a heating and cooling mechanism. The importance of radiative diffusion has first be pointed out
by \cite{2006A&A...459L..17P}, and recent papers quote approximate formulas for viscous and diffusive disks
(\cite{2010ApJ...723.1393M,2010MNRAS.tmp.1436P}).
To demonstrate the effect for two-dimensional flat disks, we show results for a planet-disk simulation where
we include viscous heating, local radiative cooling
as well diffusive radiative transport in the disk's plane.  The energy equation then reads
\begin{equation}
\label{eq:kley-energy}
 \frac{\partial \Sigma c_{\rm v} T}{\partial t} + \nabla \cdot (\Sigma c_{\rm v} T {\bf u} )
       =  \,  - p  \nabla \cdot {\bf u}
       \,  +D  - Q
      \,  - 2 H \nabla \cdot \vec{F}
\end{equation}
where $\Sigma$ is the surface density, $T$ the midplane temperature, $p$ the pressure, $D$ the viscous dissipation,
$Q$ the radiative cooling and $\vec{F}$ the radiative flux in the midplane. Models where the various contributions 
on the right hand side of eq.~(\ref{eq:kley-energy}) were selectively switched off and on, have been constructed 
by \cite{2008A&A...487L...9K}.  

\begin{figure}[ht]
\begin{center}
 \includegraphics[width=0.40\textwidth]{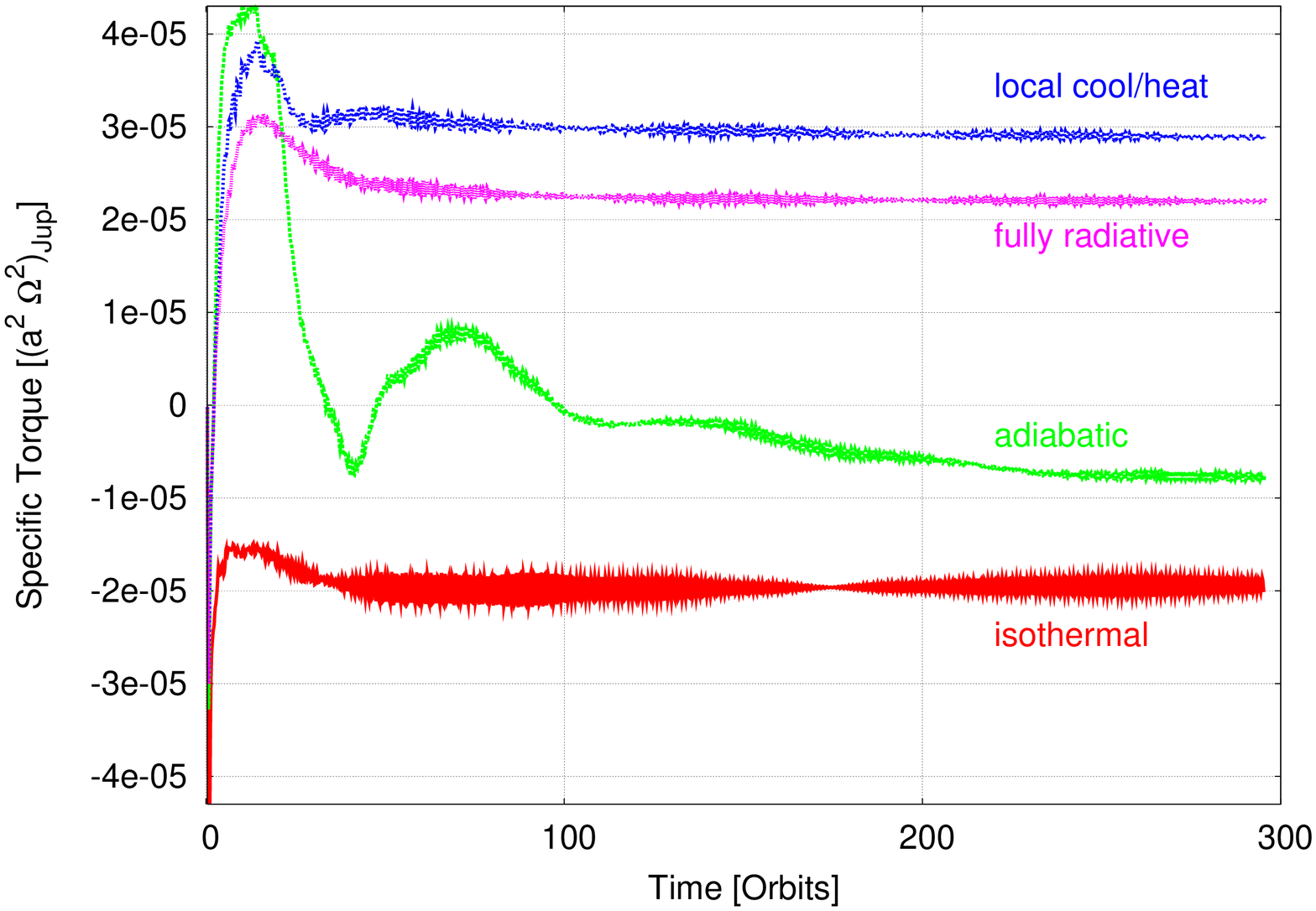} 
$~~~$
 \includegraphics[width=0.40\textwidth]{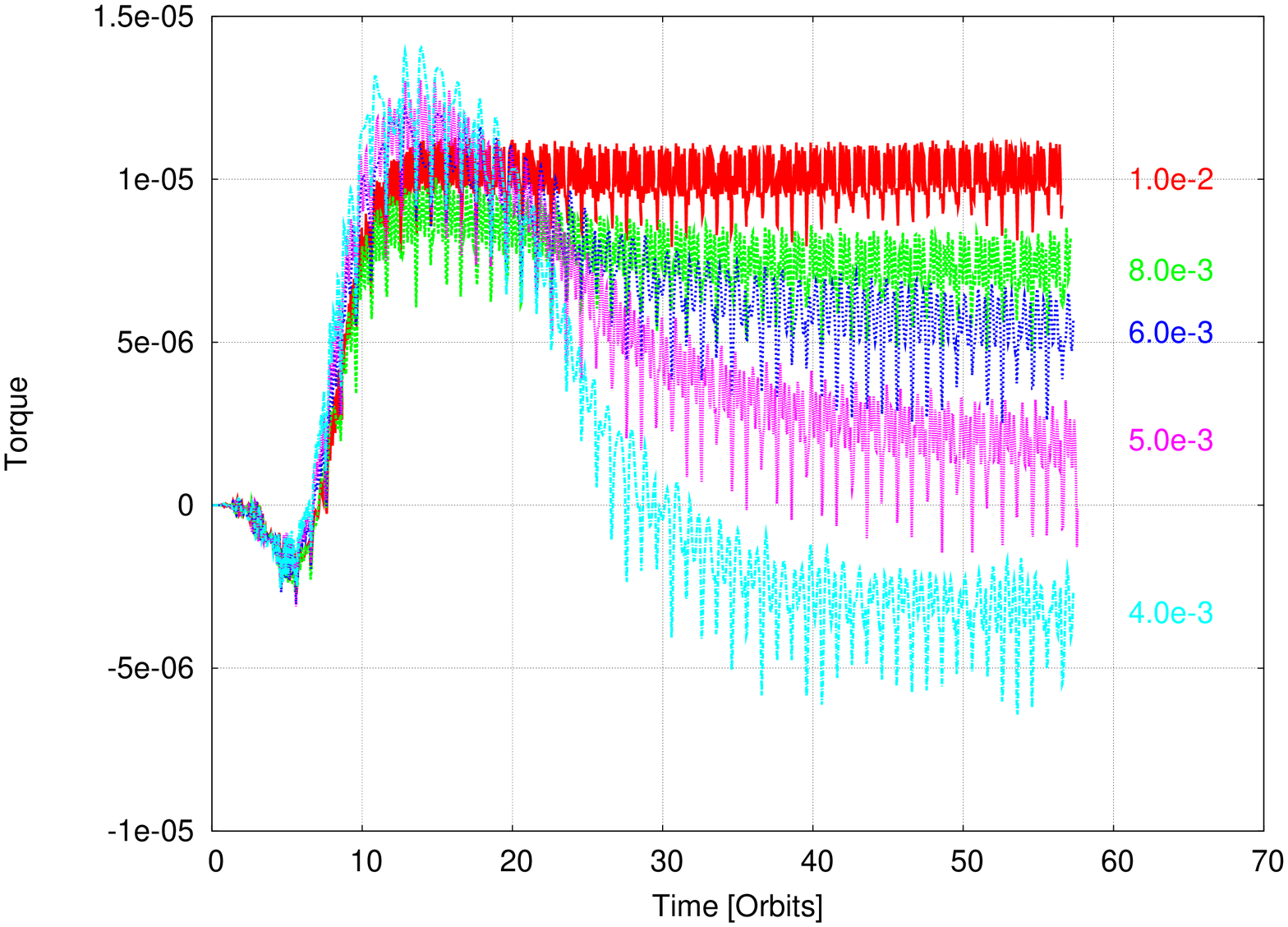} 
 \caption{
  Time evolution of the torque acting on a 20 $M_{earth}$ planet embedded in a disk
  with 0.01 solar masses with a radial range of 0.4 to 2.5 $r_{Jup}$. 
  {\bf Left}: Simulations using a constant value of the viscosity.
  From bottom to top the curves
 indicate simulations where {\it i}) no energy equation (isothermal), 
  {\it ii}) only the first term on the rhs. of eq.~(\ref{eq:kley-energy}) (adiabatic), 
  {\it iii}) all terms on the rhs. (fully radiative), and
  {\it iv}) all but the last term on the rhs. (heating/cooling), have been used. (after Kley \& Crida, 2008)
  {\bf Right}: The influence of viscosity on the resulting torque for fully radiative simulations using an
    alpha type viscosity.
    In the simulations the dissipation has been kept fixed, varying only the value of $\alpha$ in the momentum
    equation. 
}
   \label{fig:kley-torque}
\end{center}
\end{figure}
The effect of this procedure on the resulting torque is shown in the left panel of Fig.\ref{fig:kley-torque}.
The basis for all the models is the same equilibrium disk model constructed using all terms on the
rhs. of eq.~(\ref{eq:kley-energy}) and no planet. Embedding a planet of $20 M_{earth}$ yields in the long run
positive torques only for the radiative disks, where the maximum effect is given when only viscous heating and local radiative
cooling are considered. The inclusion of diffusion in the disk midplane yield a slightly reduced torque.
Since the initial state consists of a non-vanishing negative radial entropy gradient, the adiabatic model shows a positive
torque during the first 20 orbits directly after insertion of the planet. 
However, in the long run the torque becomes negative as in the isothermal case because the material within
the horseshoe region is mixed thoroughly, wiping out the entropy gradient.
Hence, the positive corotation torque is obliviated and the negative Lindblad contribution dominates.
The adiabatic case does not approach the isothermal result because the corresponding sounds speeds are different.
The right hand panel of Fig.\ref{fig:kley-torque} demonstrates that a non-vanishing value of the viscosity is
necessary to maintain torque desaturation. These models have used an $\alpha$-type viscosity in contrast to the
results displayed on the left side. To compare results directly, it has been ensured that the thermal state of
the models is identical despite the different value of $\alpha$ used in the angular momentum equation.

\begin{figure}[ht]
\begin{center}
\includegraphics[width=0.45\textwidth]{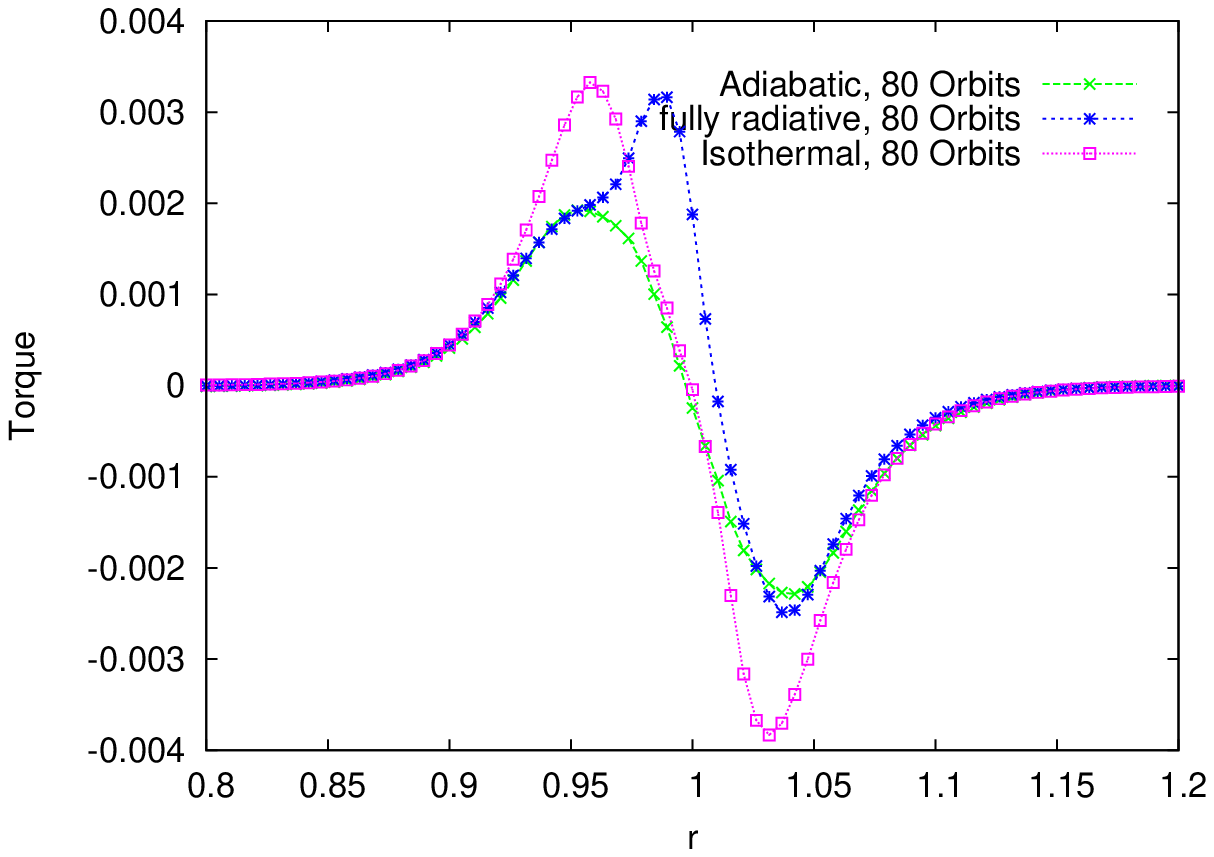} 
\includegraphics[width=0.40\textwidth]{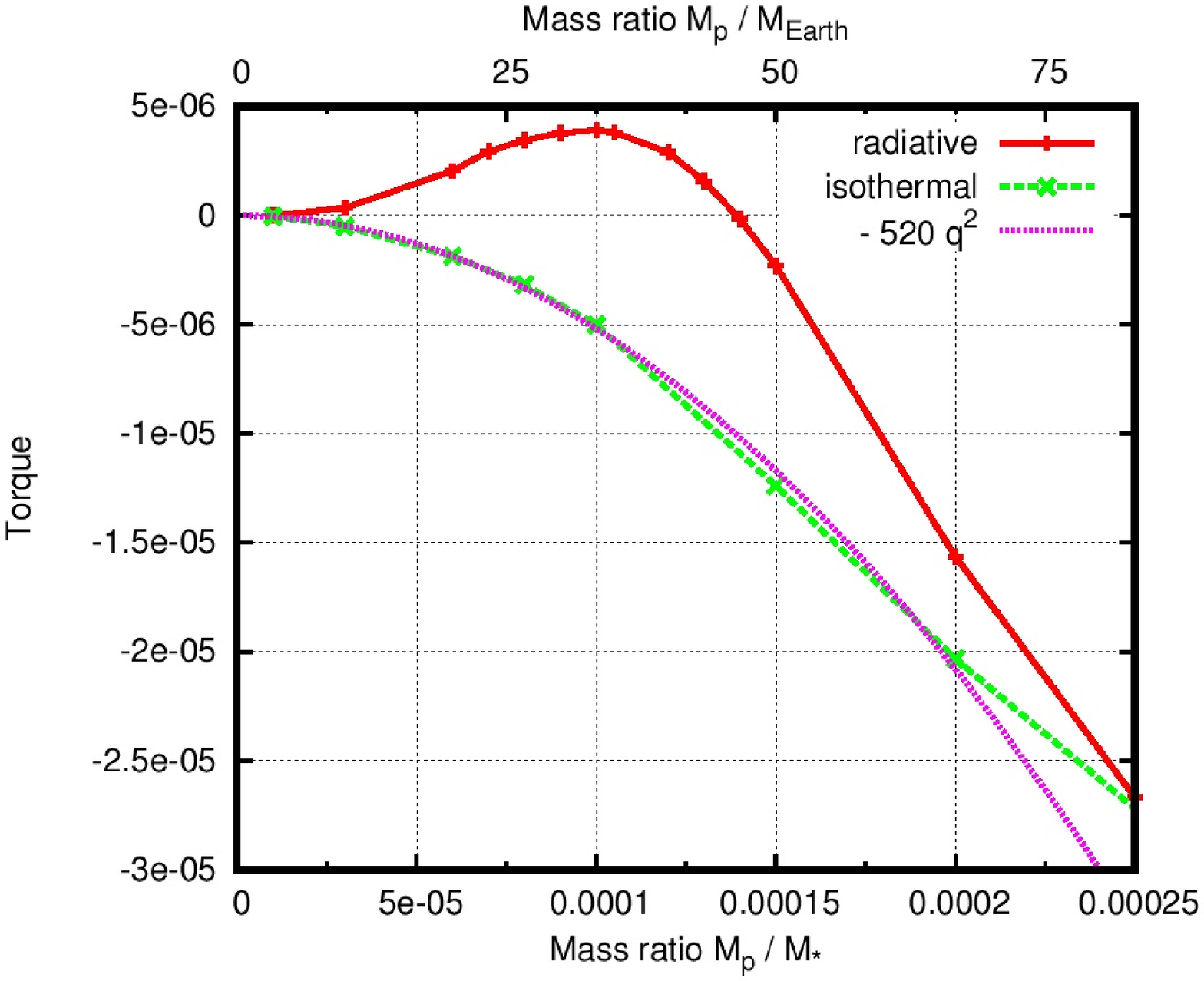} 
 \caption{{\bf Left}: Radial torque density, $\Gamma(r)$, for fully 3D radiation-hydrodynamical simulations with an embedded
  $20 M_{earth}$ planet. The data are shown after equilibrium has been reached.
 Different assumptions for the thermodynamical state are made. (after Kley et al., 2009) 
 {\bf Right}: The total torque acting on the planet for a 2D fully radiative simulation for a variety of planet
 masses.
   (after Kley \& Crida, 2008)
}
   \label{fig:kley-rad-torque}
\end{center}
\end{figure}

\subsection{Fully three-dimensional radiative disks}
The previous 2D cases have been repeated using full 3D radiative disk simulations with an identical physical setup. 
These clearly confirmed the existence of outward migration in the presence of radiative diffusion.
To compare isothermal, adiabatic and fully radiative simulations, all start from the same
initial state which corresponds to an equilibrium of the fully radiative case including viscous heating and
radiative diffusion. 
The spatial origin of the torques can be analyzed using for example the radial torque density $\Gamma(r)$ which
is defined through $\Gamma_{tot} = \int \Gamma(r) dr$, where $T_{tot}$ is the total torque acting on the planet.
A plot of the radial torque distribution (Fig.~\ref{fig:kley-rad-torque}, left panel) shows that at $t=80$ the 
corotation torques have saturated in the isothermal and adiabatic case, and only the Lindblad contributions remain.
Obviously, the net effect is the sum of two contributions that have opposite
sign and are of comparable magnitude. The negative part of the outer spiral arm has 
a slightly larger amplitude than the inner contribution. Again, the isothermal and adiabatic torques
differ due to the different sound speed.
The fully radiative case agrees for radii larger than $a_p$ with the adiabatic model while there exists
a well pronounced torque maximum just inside of the planet. This contribution is responsible for the torque
reversal. The right panel of Fig.~\ref{fig:kley-rad-torque} shows that the strength of this positive corotation
effect also scales with the square of the planet mass up to about 20 to 25 $M_{earth}$. Beyond this mass,
gap opening begins and only the Lindblad torques remain, and above 40 $M_{earth}$ planets begin to
migrate inwards again. In the full 3D simulations the results are qualitatively the same, the averaging procedure,
that is necessary in 2D, leads to some quantitative differences \cite{2009A&A...506..971K}. Interestingly, the full 3D results
show even a stronger effect. New population synthesis models based on the modified migration rates indicate better agreement
with the observational data set (eg. Mordasini, this volume). 

\section{Eccentricity and inclination}
In addition to a change in semi-major axis, planet-disk interaction will modify the planetary
eccentricity ($e$) and inclination ($i$) as well. Extending previous isothermal
studies by \cite{2007A&A...473..329C}, fully 3D radiative disk simulations have been performed recently
to study the evolution $e$ and $i$.
The results indicate that both are damped for all planetary masses (\cite{2010A&A...523A..30B}, and this volume).
For small values of $e$ and $i$ that
are below about $2 H/r$ the damping occurs exponentially on timescales comparable to the
linear estimates by \cite{2004ApJ...602..388T}. For larger values, damping is slowed down and follows approximately $\dot{e} \propto e^{-2}$,
and for the damping of $i$ an identical relation holds, surprisingly.
Interestingly, the presence of outward migration is coupled to the magnitude of $e$ and $i$. 
Outward migration only occurs for eccentricities smaller than about 0.02, and inclinations below about $4^o$.
This reduction is due to the fact that for non-circular orbits the flow structure
in the corotation region becomes strongly time dependent and no stationary corotation torque can develop.
More details about the evolution of inclined and eccentric planets in 3D disks are given in \cite{2010A&A...523A..30B},
and Bitsch \& Kley (this volume). 

\begin{figure}[ht]
\begin{center}
\includegraphics[width=0.38\textwidth,angle=-90]{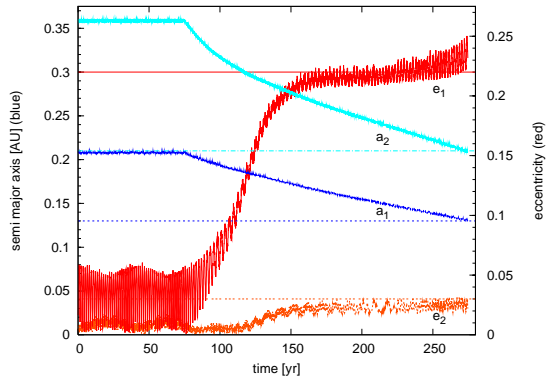} 
 \caption{Evolution of the semi-major axis $a$ and eccentricity $e$ of a pair of planets with
  physical parameter resembling the observed planetary system GJ~876.
  For the initial 75 years the planets are kept fixed and thereafter they are allowed 
  to migrate freely within the disk. The index $1$ refers to the inner and $2$ to the outer
  planet. The horizontal lines refer to the observed values of GJ~876. The evolution has been
  stopped after the planets have reached the observed distances from the star.
   (after \cite{2008A&A...483..325C})
}
   \label{fig:kley-gj876}
\end{center}
\end{figure}

\section{Resonant systems}
The mere existence of resonant planetary systems is a strong indication that dissipative mechanisms
changing the semi-major axis of planets must have operated, as the likelihood of forming planets
in these configurations in situ or later through scattering processes seems to be small.
On the other hand, convergent differential migration of a pair of planets will lead under
quite general conditions to capture into resonance.
The presence of an inner disk in combination with realistic inner boundaries
leads to very good agreement with the observations for the best observed system GJ~876,
as demonstrated in \cite{2008A&A...483..325C} and shown in Fig.~\ref{fig:kley-gj876}.
Through transit timing measurements the first double eclipsing system, Kepler~9, has been discovered,
and the data clearly indicate a low order mean motion resonance, probably 2:1 (\cite{2010Sci...330...51H}). 
The proximity of the planets to the star
and the near coplanarity of the system strongly hints towards a migration scenario for the
formation of the system.

\begin{figure}[ht]
\begin{center}
\includegraphics[width=0.80\textwidth]{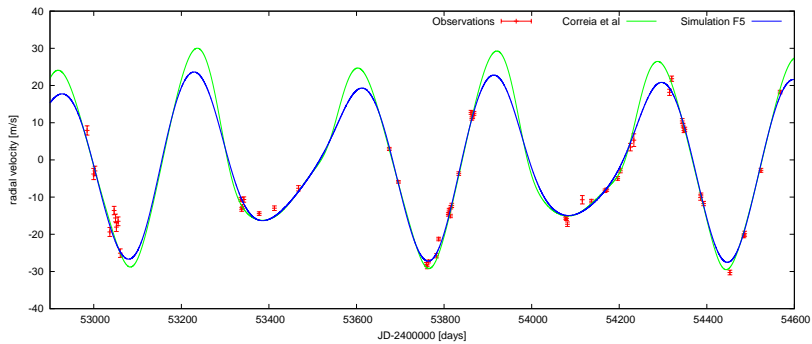} 
 \caption{Synthetic radial velocity curves HD~45364 together with observational data points.
  The light green curve is the fit as given in the discovery paper (\cite{2010A&A...510A...4R}).
  The darker blue curve has been obtained by performing full hydrodynamical simulation of the
  system. Both curves have comparable $\chi^2$ values.
}
   \label{fig:kley-hd45364}
\end{center}
\end{figure}

The importance of dynamical migration models for systems of planets is indicated by the system
HD~45364, where two planets engaged in a 3:2 resonance have been discovered by \cite{2009A&A...496..521C}.
The inferred orbital parameters for the two planets are
semi-major axes of $a_1 = 0.681$AU and $a_2=0.897$AU, and eccentricities of $e_1 = 0.168$ and $e_2=0.097$, respectively.
Fully non-linear hydrodynamical planet-disk models have been constructed for this system
by \cite{2010A&A...510A...4R}. For suitable disk parameter, the planets enter indeed 
into the 3:2 resonance through a convergent migration process. After the planets have reached their observed semi-major axis, a theoretical
RV-curve has been calculated. Surprisingly, even though the hydrodynamically obtained eccentricities ($e_1 = 0.036, e_2 = 0.017$)
are quite different, the model fits the observed data point equally well as the published data, see Fig.~\ref{fig:kley-hd45364}
and \cite{2010A&A...510A...4R}.
This pronounced dynamical difference between the two models that match the observations equally well,
can only be resolved with more observational data. Hence, the system HD~45364 is a very good example
that it is urgently necessary to have sufficient and better observational data for 
interacting multi--planet systems to obtain the orbital parameters, that are required to constrain the theoretical models.

\section{Summary}
We have shown that just by inclusion of more accurate physics, namely radiation transport, it is possible to
reduce significantly the otherwise too rapid type I migration. The effect is driven by corotation torques,
and the requirement to maintain the horseshoe torques unsaturated is the
presence of viscosity and radiative cooling (or diffusion), with timescales of the order of the libration
time of the horseshoe material near the separatrix. 
Including this effect, population synthesis models yield better agreement with the observations.
We find that eccentricity and inclination are typically damped by planet-disk interaction for all planet
masses.

Differential migration in multi-planet systems frequently result in resonant capture. Upon capture the
eccentricity of the planets is strongly increased. For continued migration the systems remain stable
only when eccentricity is damped by the disk. We have shown that the formation of the 
systems GJ~876 and HD~45364 can naturally be explained by planet-disk interaction migration scenarios.
For resonant systems where the disk action is reduced in the final stages of the planet formation process
the left over configuration may result in an unstable system. The last phase will then be dominated by
scattering processes which may pump up the planetary eccentricities and possibly their inclinations 
to the large values observed.

\end{document}